# System modeling of a health issue: the case of preterm birth in Ohio

Authors: Alireza Ebrahimvandi, Niyousha Hosseinichimeh

## Abstract

Preterm birth rate (PBR) stands out as a major public health concern in the U.S. However, effective policies for mitigating the problem is largely unknown. The complexities of the problem raise critical questions: Why is PBR increasing despite the massive investment for reducing it? What policies can decrease it? To address these questions, we develop a causal loop diagram to investigate mechanisms underlying high preterm rate in a community. Our boundary is broad and includes medical and education systems, as well as living conditions such as crime rate and housing price. Then, we built a simulation model and divided the population into two groups based on their chance of delivering a preterm baby. We calibrated the model using the historical data of a case study—Cuyahoga, Ohio—from 1995 to 2017. Prior studies mostly applied reductionist approaches to determine factors associated with high preterm rate at the individual level. Our simulation model examines the reciprocal influences of multiple factors and investigates the effect of different resource allocation scenarios on the PBR. Results show that, in the case of Cuyahoga county with one of the highest rates of PBR in the U.S., estimated preterm birth rates will not be lower than the rates of 1995 during the next five years.

## Background

The U.S. preterm birth rate (PBR) —the percentage of births before 37 completed weeks of gestation— is a national health problem. In the latest global ranking of PBR, the U.S. ranked 131st out of 184 countries [1]. In addition, the lowest PBR of the 21st century in the U.S., which happened in 2014 is still 2% higher than the rate of 9.4% in 1981 [2, 3].

State of Ohio faces constantly higher rates of preterm birth despite the investments and efforts to reduce the PBR as one of the most important indicators of the well-being of society [4]. This problem is dire in some of its counties like Cuyahoga. Figure 1 shows the PBR in the U.S., Ohio, and Cuyahoga County from 1995 to 2017. Ohioans have access to one of the best health care systems and they are one of the richest states. Ohio's PBR ranks 32nd in the nation and this rate has increased from 11.21 to 11.94 between 1995 and 2017. This unfavorable PBR ranking is despite a strong ranking of 12th in health care access and 7th in gross domestic product among 50 states in the U.S. [5, 6]. Cuyahoga County has one of the highest rates of PBR in Ohio and the U.S.

The PBR problem in Ohio is more troubling in terms of racial disparity. African Americans have higher rates of PBR among all races. African American mothers are also 30 percent more likely to deliver a preterm baby in Ohio compared to California. (African American's PBR was 13.3 and 17.7 in California and Ohio between 2007 to 2017, respectively [7].)

Preterm birth imposes long-term economic, educational, and social costs to society; therefore, it is a significant public health concern. The Institute of Medicine reported that the annual cost of preterm birth in the U.S. was $26.2 billion in 2005 [8]. Further, many babies born preterm suffer from lifelong deficiencies [9, 10]. Preterm birth can also both be the cause and result of social inequality [11, 12]. For instance, a preterm born person can have less income or educational attainment due to lack of the

cognitive development and on the other hand, the underprivileged are at higher risk of giving of preterm delivery [12, 13].

Many risk factors are associated with higher risks of preterm birth. The most important biomarker predictors of preterm birth are a history of preterm birth, genetic markers like fetal fibronectin, and pregnancy characteristics like the number of fetuses [14]. However, genetics can explain only about 23% of preterm births [15]. Also, the exact biological pathways causing preterm birth are still unknown [14], Iams, Romero [16] summarized the possible interventions at different stages of pregnancy and for different populations that may reduce the odds of preterm birth. The paper then discusses that the focus of policymakers and healthcare systems have been on the primary interventions that target women during their pregnancy.

The stagnating rates of preterm birth suggest that policymakers may need to change their focus from medical interventions during pregnancy to the time where mothers become vulnerable to the risk of preterm birth. A growing body of literature emphasizes the importance of life-course perspective on the birth outcomes. The life-course perspective considers intergenerational socio-economic factors to explain the problem of preterm birth [12, 17]. In the next section, we use the life-course theory as well as quantitative and qualitative data to build an SD simulation model of the preterm birth in Cuyahoga County. We used Kim and Andersen [18] to transform qualitative data to feedback loops.

The PBR trend suggests that the current approach of targeting this public health concern has been ineffective. Preterm birth is a complex problem that is more of a chronic process than an acute one [19]. The authors include the life-course perspective theories and employ the tools that are capable of considering these complexities over time. System dynamics (SD) is the approach that we use for analyzing this problem. The SD is a methodology for understanding and evaluating the nonlinear behavior of complex problems and feedback loops over time [20]. This methodology was used in analyzing various public health issues such as depression, obesity, infant mortality, diabetes, and post-traumatic stress disorder to provide insights to policymakers in understanding the problem and providing solutions [21-27].

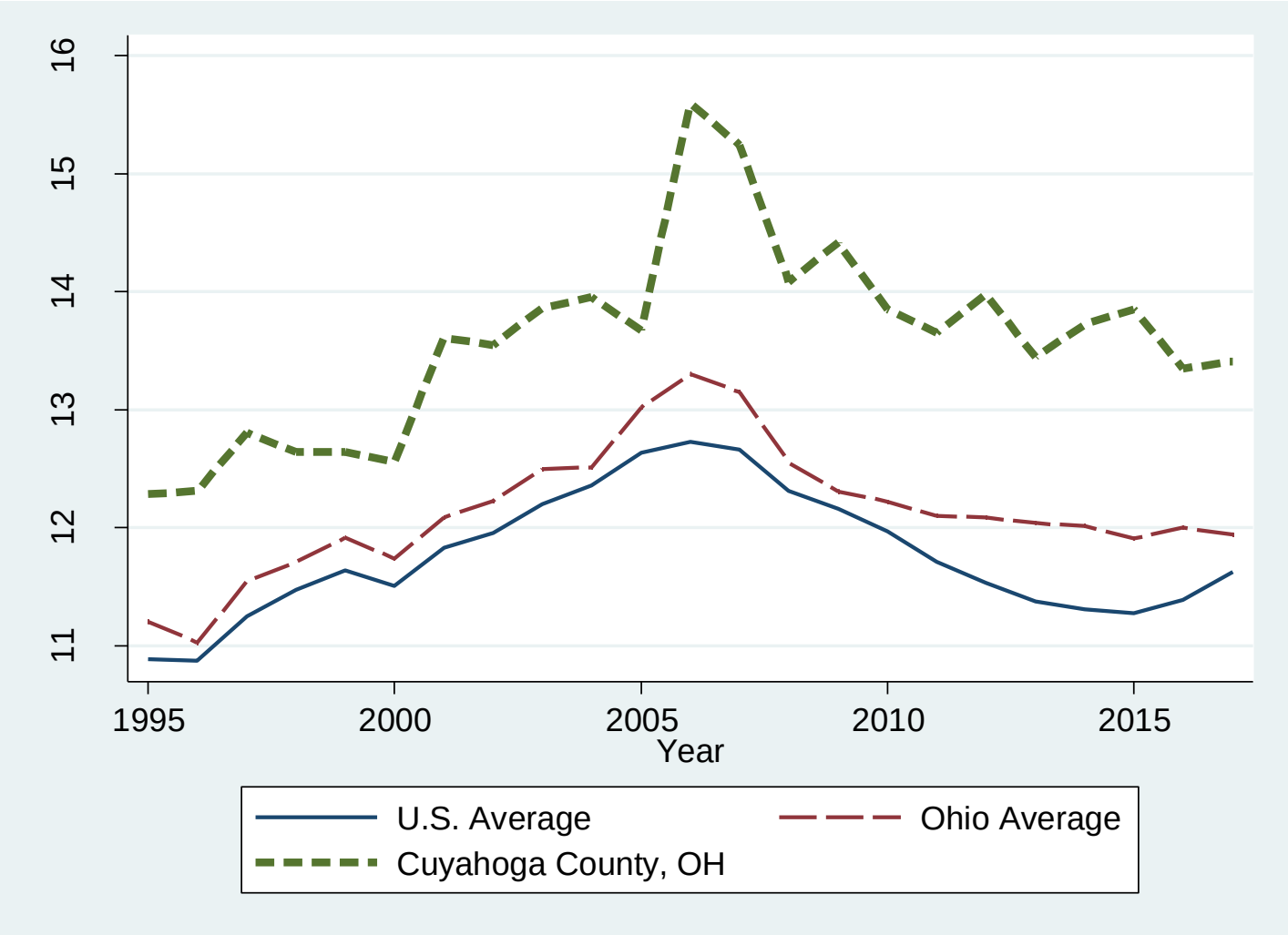

*Figure 1  Preterm birth trend in the U.S., Ohio, and Cuyahoga County*

## Model and data

Theory of life course shifts the focus from health services during pregnancy to when the biological and physiological characteristics of mothers are shaped. This theory suggests that exposure to acute and chronic stress over time erodes the ability of the immune system to work effectively [28]. There is compelling research that shows maternal stress and corticotropin-releasing hormone (CRH) can cause preterm parturition in two possible ways. First, chronically increased CRH results in suppression of the immune system [12]. Second, the elevated CRH levels may impair counter-regulation of hypothalamic-pituitary-adrenal levels, which leads to immune-inflammatory dysregulations. This phenomenon, elevated CRH, increases the relative risk of a spontaneous preterm birth more than three-fold [29].

Mothers who are exposed to stress during their lifetime are more vulnerable because they have higher levels of allostatic load. Allostatic load is defined as the cumulative physiological toll that accumulates through financial distress, illness or injury, exposure to environmental hazards, or risky behaviors throughout life [12]. Furthermore, living in an economically strained household has a great impact on the stress level of its members. Vulnerable people are at a higher risk of preterm birth [30]. This is because the vulnerable have a higher propensity for engaging in unhealthy behaviors like bad diet, smoking, and drinking that are also risk factors of preterm birth [12].

Education has a unique dimension in forming the social and health status. Educational attainment is the main bridge between the status of one generation and the next. It also functions as the “main avenue of upward mobility” [31]. Educated people are healthier not because they can afford better or more health services, but because they can buy themselves out of privation. Education increases the sense of control, which eventually leads to a healthy and less stressful lifestyle. Therefore, it is necessary for society to invest in education as the best tool available for changing the social status of children living in vulnerable communities. This transition is costly and might take years to show results.

### Causal loop diagram

**R1- Vulnerable population and less resources on education**; As the share of vulnerable population increases in a community, per capita resources decline. Both families and governments have less resources to invest in education and safety, which in the long run leads to even more vulnerable populations. Figure 3 shows this causal effect in a loop.

Although the average income in Cuyahoga County increased between 1995 to 2016, the percentage of people in poverty has also increased. Financial distress increases the allostatic load and makes the population more vulnerable to the risk of preterm birth. The state of Ohio received less in taxes from its residents compared to its neighboring states in 2016. Ohio residents paid an average of $2,471 per capita in state taxes in 2016, while residents in Indiana, Michigan, and Pennsylvania paid higher taxes at $2,652, $2,763, and $2,925 per person, respectively [32].

**R2- Short-term investment on medical care**; In a society with increasing vulnerable populations, per capita resources decline. This then reduces the financial flexibility of both families and governments to invest in healthcare. With less investment in healthcare, the average health outcomes get worse with a delay and subsequently birth outcomes deteriorate. Furthermore, the vulnerable population requires more medical attention, because they have more risky behavior like smoking, drinking, and drug abuse that are also risk factors of preterm birth [12].

The poor health outcomes like high preterm birth or infant mortality rate increase the gap between the actual outcomes and the desired outcomes. The higher gap pressures governments and families to invest more in Medicaid and medical care, respectively. This increase in healthcare spending improves the health outcomes in the short term, but it drives away the resources from investment in schools and safety. Less investment in education and safety increases the vulnerable population in the long run.

In Ohio, the highest preterm birth rate occurred in 2006 (see Figure 1). Consequently, the Ohio Perinatal Quality Collaborative (OPQC) was formed in 2007 to reduce the PBR [33]. Ohio has also expanded the short cervix screening to all women with the hope that this intervention would reduce the PBR rates in 2007 after the spike in 2006 [34].

**R3- Wealthier is healthier**; A wealthier community has more resources to invest in healthcare. As the healthcare outcomes improve, the community can invest a larger portion of its resources in education and safety. This will decrease the vulnerable population in the long run in two ways. First, investing in the education of children in the community makes them less vulnerable in the long run. Second, funding schools makes the neighborhood more attractive and increases the inflow of people that are in better financial situations. As a result, the share of the vulnerable population declines and the community has even more resources to invest.

**R4- Fight or flight**; The crime rate is higher in communities with more vulnerable populations. As the crime rate increases, the attractiveness of the community goes down and those in better financial and health condition relocate to a safer neighborhood. Humans respond to perceived threats with a primitive, biological, fight-or-flight decision. This decreases the demand for real estate in the neighborhood and decreases the house price index. As a result, low-income families that have limited housing options may have no other choice but to move to these neighborhoods. This further increases the share of vulnerable populations in the community.

According to Zack Reed, the Ward 2 Councilman of Cleveland, Cuyahoga, they lost 26 percent of their population because of two fundamental issues: crime and schools [18, 35]. Cuyahoga County had relatively higher rates of crimes compared to the U.S. average (Figure 2.a). Violent crime includes murder, rape, robbery, and aggravated assault.

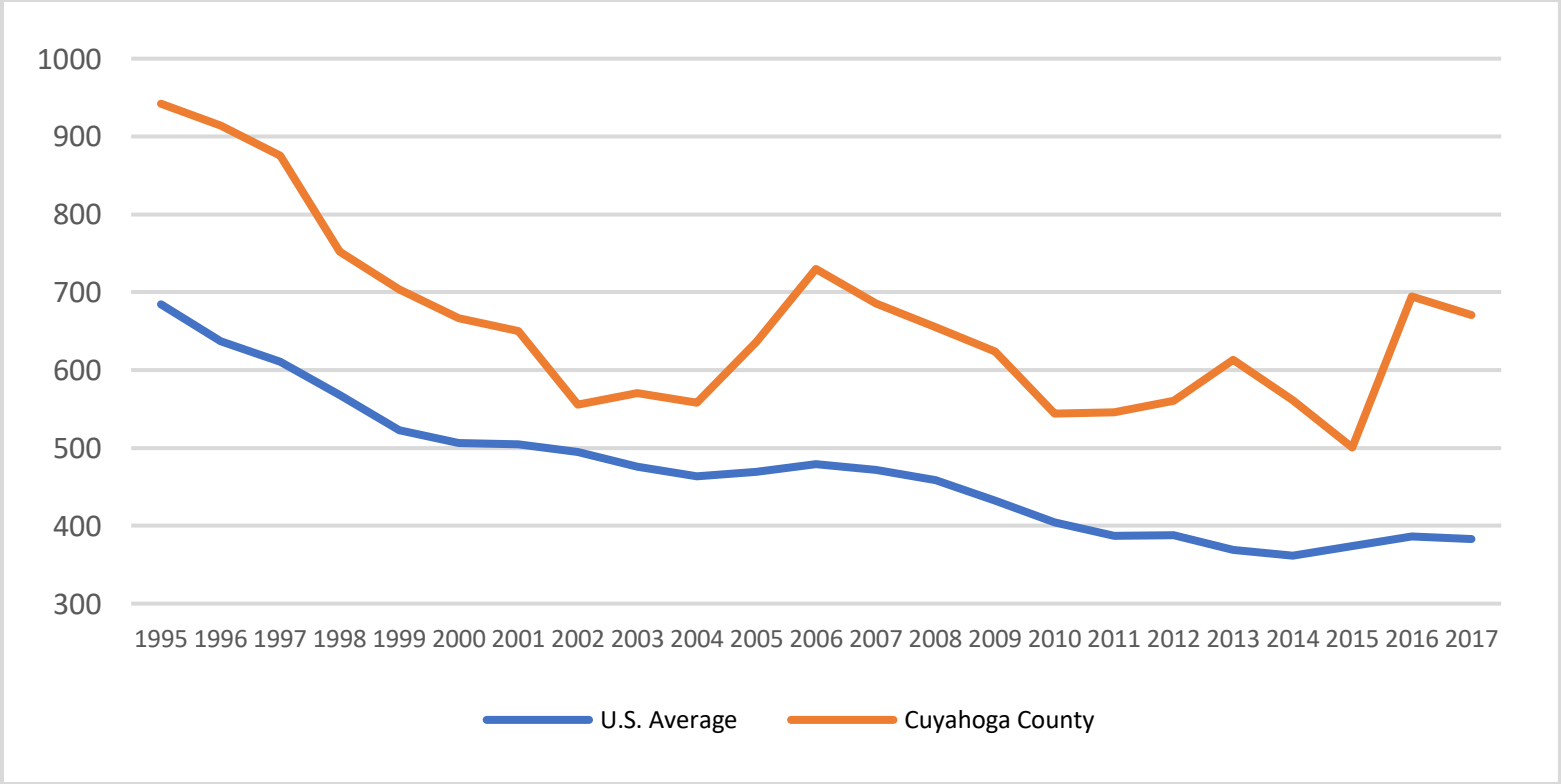

*Figure 2. The violent crime rate in the U.S. and Cuyahoga County*

**R5- Attractiveness of schools and safety**; Communities with fewer resources have less financial flexibility in allocating them to education, safety, and correction. Lower investment in education and safety makes the neighborhood a less attractive place to live. The decline in demand for houses reduces the property value. Therefore, the neighborhood becomes more affordable for vulnerable populations. This keeps the share of vulnerable populations high in the neighborhood and as a result, the financial resources per person remain low.

**B1- Medical care**; When a community has poor health outcomes, the preterm birth rate increases. This then increases the pressure to allocate more resources to healthcare, especially Medicaid, to improve the birth outcomes. More resources for Medicaid improve access to healthcare and subsequently health outcomes.

Ohio Medicaid covered pregnant mothers who were living in a family with less than 150% of the federal poverty level (FPL) in 2003 and this threshold has significantly increased to 205% in 2018. The ranking of Ohio in terms of Medicaid expansion has also improved. Ohio has decreased its ranking from 37$^{th}$ in 2003 to 25$^{th}$ in 2018 [36]. This improvement shows that Ohio is covering not only more of the vulnerable population compared to 2003, but it has also improved the health coverage faster than the national average.

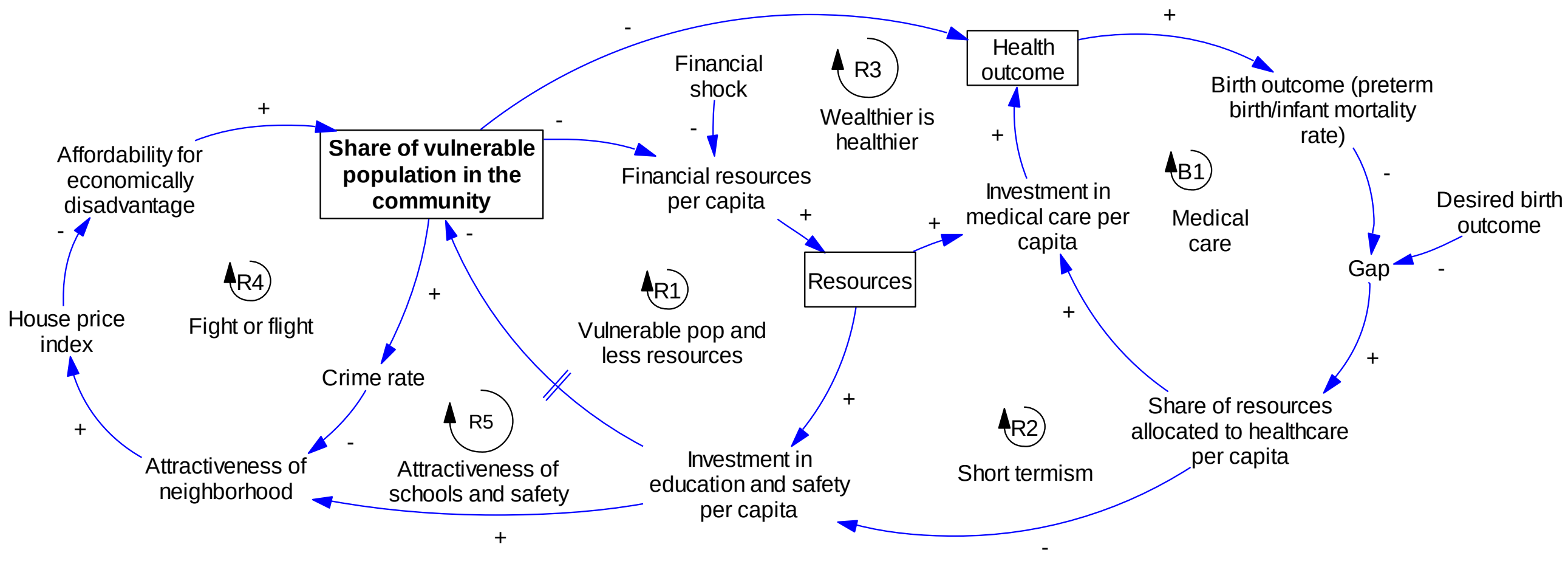

*Figure 3 Causal loop diagram*

In addition to the loops, we also consider two exogenous variables in our CLD. The first is a financial shock and the second is the desired rate of preterm birth. There are times that the community encounters an exogenous financial shock. A system with a larger rainy-day fund has more resources to handle the shock without substantial reduction of its investment in education or healthcare. Ohio and its counties experienced many financial shocks in the past two decades especially after auto manufacturing companies like Ford laid off 35,000 workers in 2001.

The desired value of preterm birth rate changes the priorities in resource allocation. For example, we have observed that Governor Kasich is more sensitive to the birth outcomes so he is allocating more resources to improve these outcomes.

Resource misallocation can affect people’s social status and health outcomes in different ways. The problem becomes more noticeable at the time of an infant’s birth because both the mother and infant are at one of the highest vulnerability points in their lives. The priorities in the allocation of resources in the long-run might seem obvious after framing the problem in a causal diagram and looking at the problem at the system level. Although under pressure, policymakers might not make decisions that are optimal in the long run and might instead just choose an approach that postpones the problem to another time. These approaches mainly mitigate problems that are in the critical stage and the solutions are effective in the short run [37].

When the outcomes of pregnancy become undesirable, expanding Medicaid might be one of the only options that policymakers have. Policymakers can observe the effect of their investment in medical care in a relatively short period of time, despite the per capita cost of this care, compared to social programs like education. Even though social programs benefit everyone in the community, there are long delays between the investment and results. Therefore, it is hard to choose the right balance in allocating resources when the results of these allocations are not immediately manifested. Thus, it is useful to build a simulation model that can help policymakers in their decision-making process.

Table 1 shows how Ohio’s budget from 2010 to 2015 is allocated between different categories. The resources that are allocated to K-12 education decreased from 20.2% to 16.8% of the total budget. The same trend in the decreased allocation of resources also occurred in higher education. On the other hand, the share of Medicaid funds increased from 21.3% to 37.4% of the budget.

*Table 1 Ohio’s spending from 2010 to 2015 in each category by percentage*

| Year | K-12 Education | Higher Education | Medicaid | Corrections | Other |
|---|---|---|---|---|---|
| 2010 | 20.2% | 4.9% | 21.3% | 3.4% | 50.2% |
| 2011 | 17.7% | 4.6% | 23.2% | 3.2% | 51.3% |
| 2012 | 20.6% | 4.2% | 24.4% | 3.1% | 47.7% |
| 2013 | 17.0% | 4.3% | 29.2% | 3.2% | 46.3% |
| 2014 | 16.8% | 4.2% | 35.8% | 3.0% | 40.2% |
| 2015 | 16.8% | 4.1% | 37.4% | 2.9% | 38.8% |

Source: National Association of State Budget Officers
Note: "Other" expenditure includes public assistance, transportation, Children's Health Insurance Program (CHIP), state police, economic development, employer contributions to pensions and health programs.

During the same time period, neighboring states like Indiana, Michigan, and Pennsylvania spent a higher percentage of their budget on K-12 education and a lower percentage on Medicaid. In 2015, the share of budget expenditure of Ohio, Indiana, Michigan, and Pennsylvania on K-12 education was 16.8%, 30.0%, 25.2%, and 18.5% and the share of spending on Medicaid was 37.4%, 31.2%, 30.2%, and 37%, respectively.

## Formulation

To build the simulation, we divided the model into three sectors: population, resources, and crime.

### Sector one: Population

The model includes two stocks for population: the stocks of low allostatic load (LAL) and vulnerable population. The flow of the stocks can happen in three ways: birth, death, and migration, either in or out of the community. In this model, we assume that a newly born baby belongs to the same social class as her parents; for example, a baby that is born in a vulnerable family stays in the same stock of population. Our model also accounts for the fact that people can enter or exit the community. This can happen for both of the population types. People may move into and out of the community due to several factors that we describe in the crime sector.

People can also move from one to another stock inside the community in two ways. First, the transition of people to the vulnerable stock happens in the case of financial shock. We assume that only a fraction of people become vulnerable as a result of financial shock. Second, the transition of people to the low allostatic load population happens through the effect of education's upward mobility. However, this mobility occurs with a delay, and only a fraction of the population makes the transition, those who have received proper education. Figure 4 shows a simplified version of the population sector. Next, we describe how we calculate the preterm birth rates in the model.

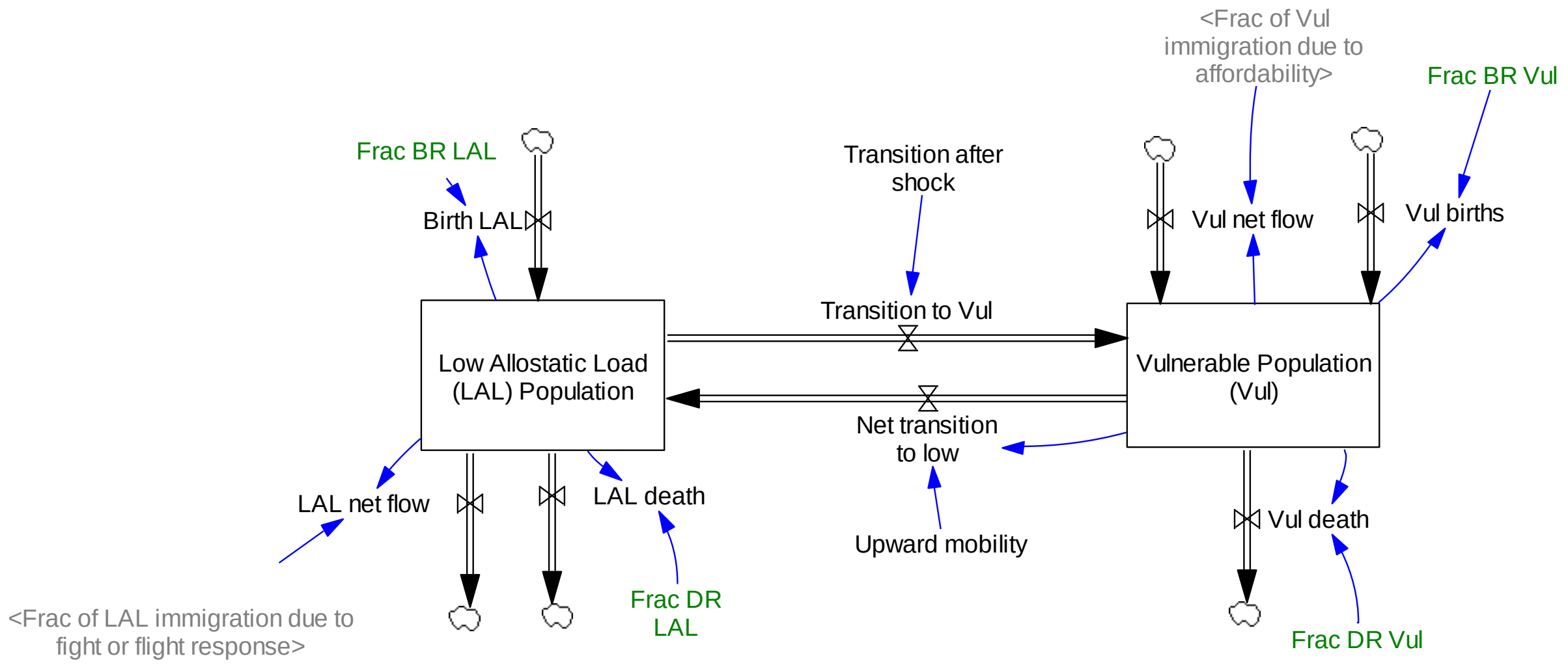

*Figure 4 Sector of population in the model*

**Preterm birth rate calculations:** To calculate the PBR, we use the weighted average of preterm birth rates between vulnerable and the LAL population. We assume that the PBR of the LAL population is equal to what the 2020 Healthy People considers as baseline [38]. This rate is 10.4%, which was the national average PBR in 2007. We also need to estimate the PBR of the vulnerable population in the model.

$$PBR = Preterm\ births/Total\ births * 100 \tag{1}$$

$$Preterm\ births = Vul\ preterm\ births + LAL\ preterm\ births \tag{2}$$

$$Vul\ preterm\ births\ =\ Vul\ preterm\ OR\ w\ prenatal\ care * Vul\ births * Preterm\ rate\ for\ LAL \tag{3}$$

$$LAL\ preterm\ births = Birth\ LAL * Preterm\ rate\ for\ LAL \tag{4}$$

To calculate the vulnerable preterm odd ratio (OR) for those who received prenatal care ($VOR$), we first need to calculate the amount of resources allocated to Medicaid and healthcare coverage for the vulnerable population, which will be described next.

**Sector two: Resources**

Government financial resources are generated through taxes that residents pay. We assume that a vulnerable person can pay a smaller amount of tax compared to a low allostatic load person. These resources are then allocated to healthcare, schools, and other categories. Healthcare allocation includes the budget of Medicaid for those who have an income less than a certain threshold. The other categories include different budget allocations including public assistance, transportation, children's health insurance program, state police, economic development, employer contributions to pensions and health programs. We also include the amount of money that the federal or local governments match, proportional to the state budget for Medicaid expansion or schools. Figure 6 shows a simplified version of the resource sector.

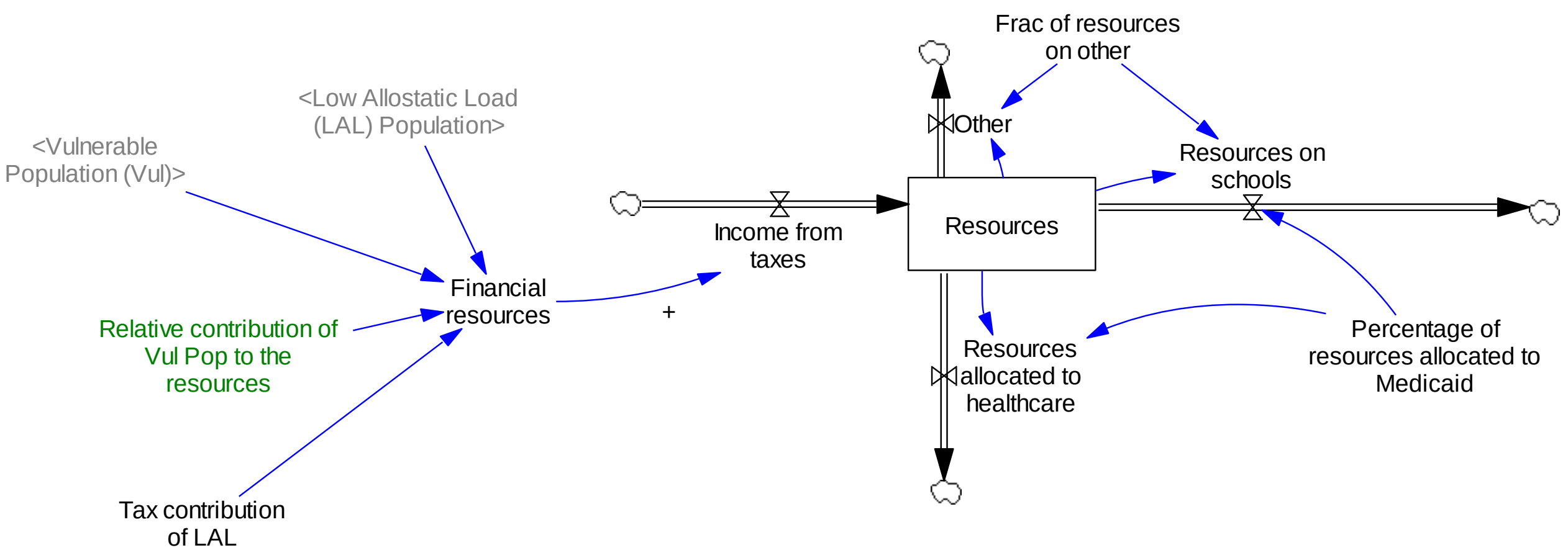

*Figure 5 Resource generation and allocation*

The variable "Percentage of resources allocated to Medicaid" is dependent on the gap between the desired rate of preterm birth and the actual preterm birth rate. As the realized gap increases, the government makes the decision to increases spending on Medicaid expansion to decrease the PBR. We used a lookup function for estimating this percentage. This increase in spending then increases the fraction of people who have health coverage. The odds ratio of preterm birth in a vulnerable population who received prenatal care ($VOR$) is calculated based on this formula:

$$VOR = Vul\ preterm\ odd\ ratio * (Medical\ care\ effect * Insured\ frac\ +\ (1 - Insured\ frac)\ ) \tag{5}$$

If none of the vulnerable mothers receives medical care, i.e. $Insured\ frac = 0$, she delivers a preterm baby at the rate of $Vul\ preterm\ odd\ ratio$. The fraction of the vulnerable population who receives health insurance ($Insured\ frac$) has a lower chance of delivering a preterm baby. This reduction is dependent on the effectiveness of medical care. The estimated effect of medical care in the reduction of the preterm birth is 86% (CI: 60-100%), which means that a pregnant mother is delivering a preterm baby with only 86% chance even after receiving medical care compared to the mothers with no prenatal care [39, 40]. The final value of the $VOR$ is the weighted average of the preterm odd ratio of vulnerable people who received care and those who did not receive any care.

**Sector three: Crime**

We compare the crime rate of the community with the national average. If the relative crime rate of the community is more than one (i.e. greater or equal than the national average), people have the perception of living in an unsafe community. Therefore, people move out of the community [31]. The price of housing in a community that has less demand than the supply drops, and this makes the neighborhood more affordable for the vulnerable population. Hence, the net immigration of the vulnerable will be more than the net outmigration of the low allostatic load population.

The detailed model formulation is in Appendix A and the model is formulated and simulated in Vensim DSS x32. The model was calibrated using historical data for preterm birth and population from 1995 to 2017.

## Data

To calculate the preterm birth rates, we used the "linked birth-infant deaths period data" files for the periods 1995-2017 [41]. We obtained the population data from the "U.S. Bureau of the Census, Resident Population in Cuyahoga County" report [42].

To simplify the measurement of the vulnerable population for each county, we consider that this figure is equal to two times the number of people below 200% of the FPL in each year. This is the threshold that Ohio used for identifying eligible pregnant mothers for Medicaid in 2018. We retrieved the number of people in poverty from the "U.S. Bureau of the Census, Estimate of People of All Ages in Poverty in Cuyahoga County" report [43].

The percentage of school-aged children in the community is 16%, and we used the percentage of people under 18 to estimate this parameter [44]. The cost of education in a public school in constant 2016-2017 dollars is $13,000 per year [45]. We derived the violent crime rates of the community from the Crime Statistics and Crime Reports published by the Ohio Department of Public Safety [46]. The national average crime rates in the U.S. is from the annual reports of crime published by the Federal Bureau of Investigation [47]. The description of other data and parameters for this model is in Appendix A.

## Results

Figure 6.a and Figure 6.b show the preterm birth rate and the population in Cuyahoga County, respectively. The blue solid lines with "1" show the simulation outputs, and the red lines with "2" show the historical data. Figure 6 shows that the model structure could generate behavior that is compatible with the historical data. It also shows that for the next five years the rates of preterm birth will not be lower than the rates of 1995.

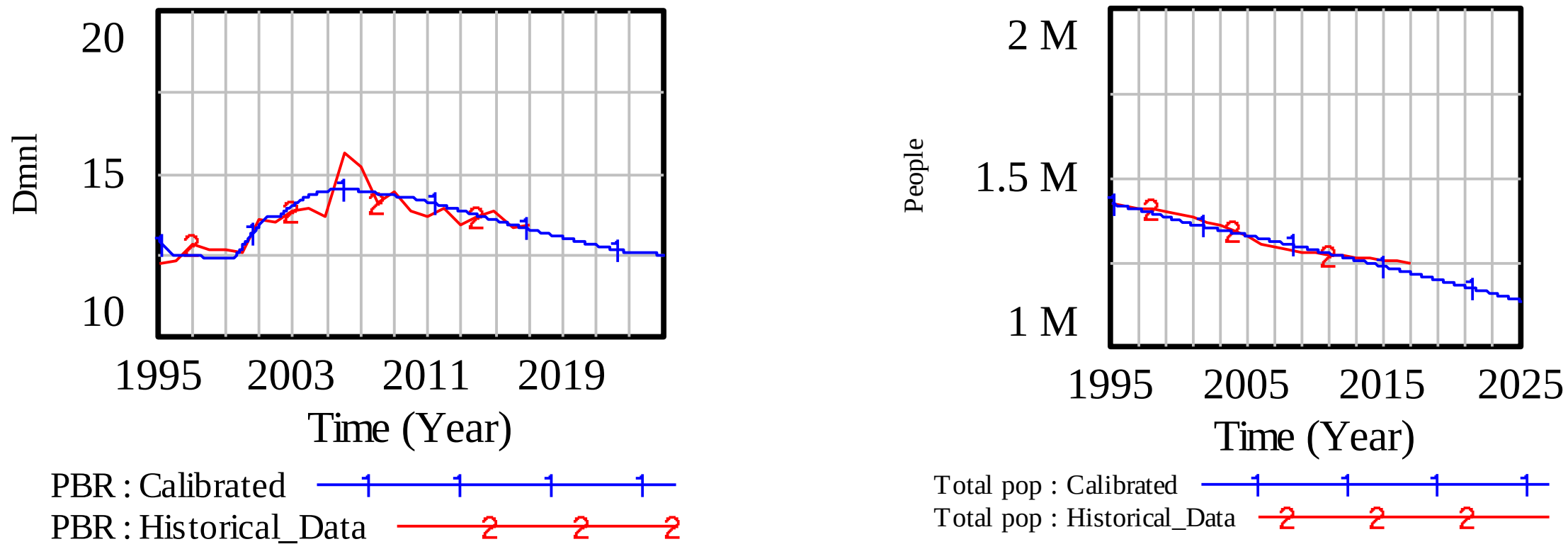

*a. Preterm birth simulated vs. historical data*

*b. The population of the county simulated vs. historical data*

*Figure 6 Calibration of the parameters*

Figure 7 shows the trend of vulnerable population in Cuyahoga county both in historical data and simulated results.

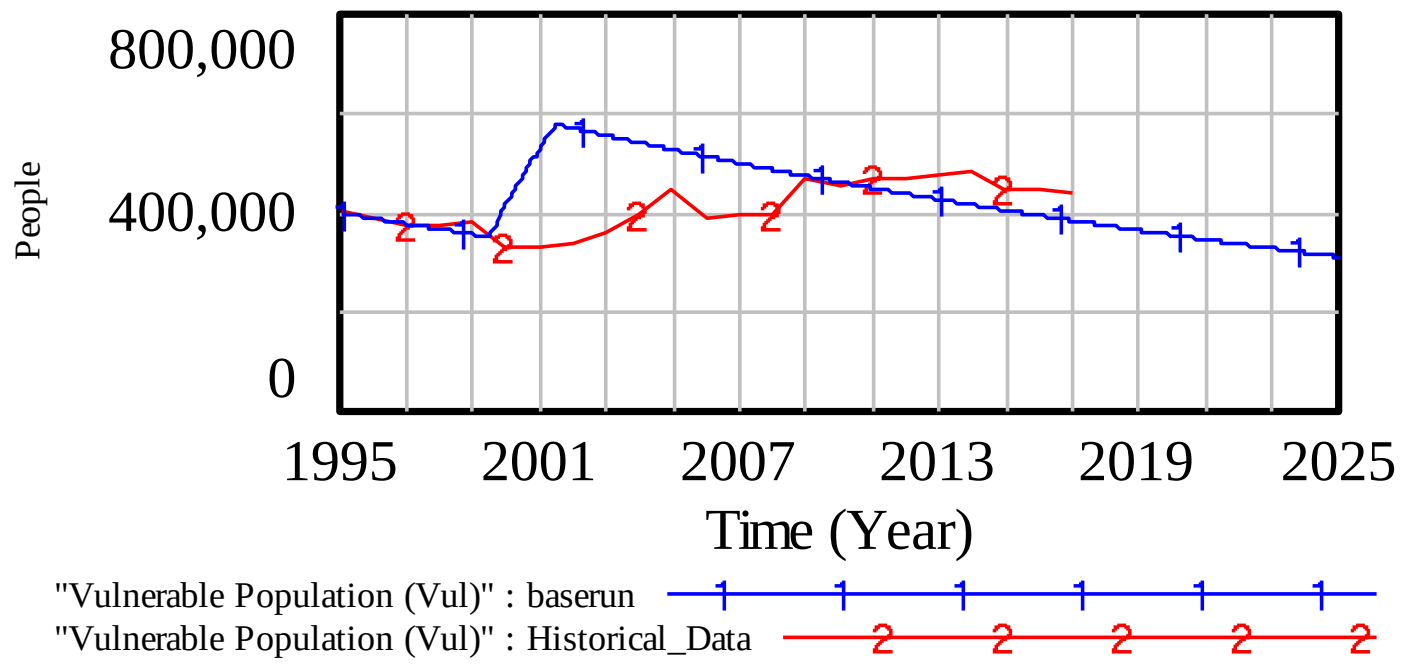

*Figure 7 Comparison of vulnerable population historical vs. simulated data*

## Policy testing

We investigate the impact of two scenarios on preterm births and resources allocated to medical care. These scenarios are compared with a base run. In the first scenario, we assume that the fraction of the vulnerable population after the financial shock increases to 22 instead of 15 percent of the base run in 2000. In the second scenario, we assume that the desired preterm birth rate is set at a better value of 9 instead of 11 in the base run.

**Scenario 1 (S1)**: Increasing the fraction of vulnerable after a financial shock;

In the base run, we considered that the Cleveland area (in Cuyahoga county) experienced a significant financial shock due to couple of auto manufacturing companies like Ford laying off 35,000 workers around the year 2000. Figure 8 shows the per capita income of Ohio residents compared to the U.S. between 1999 to 2008 [48]. This graph shows that Ohio experienced a decline in the real GDP in 2000. We include this financial shock in the base run of our model. We also considered that this shock moved only a small fraction (15%) of the people from low allostatic load to vulnerable.

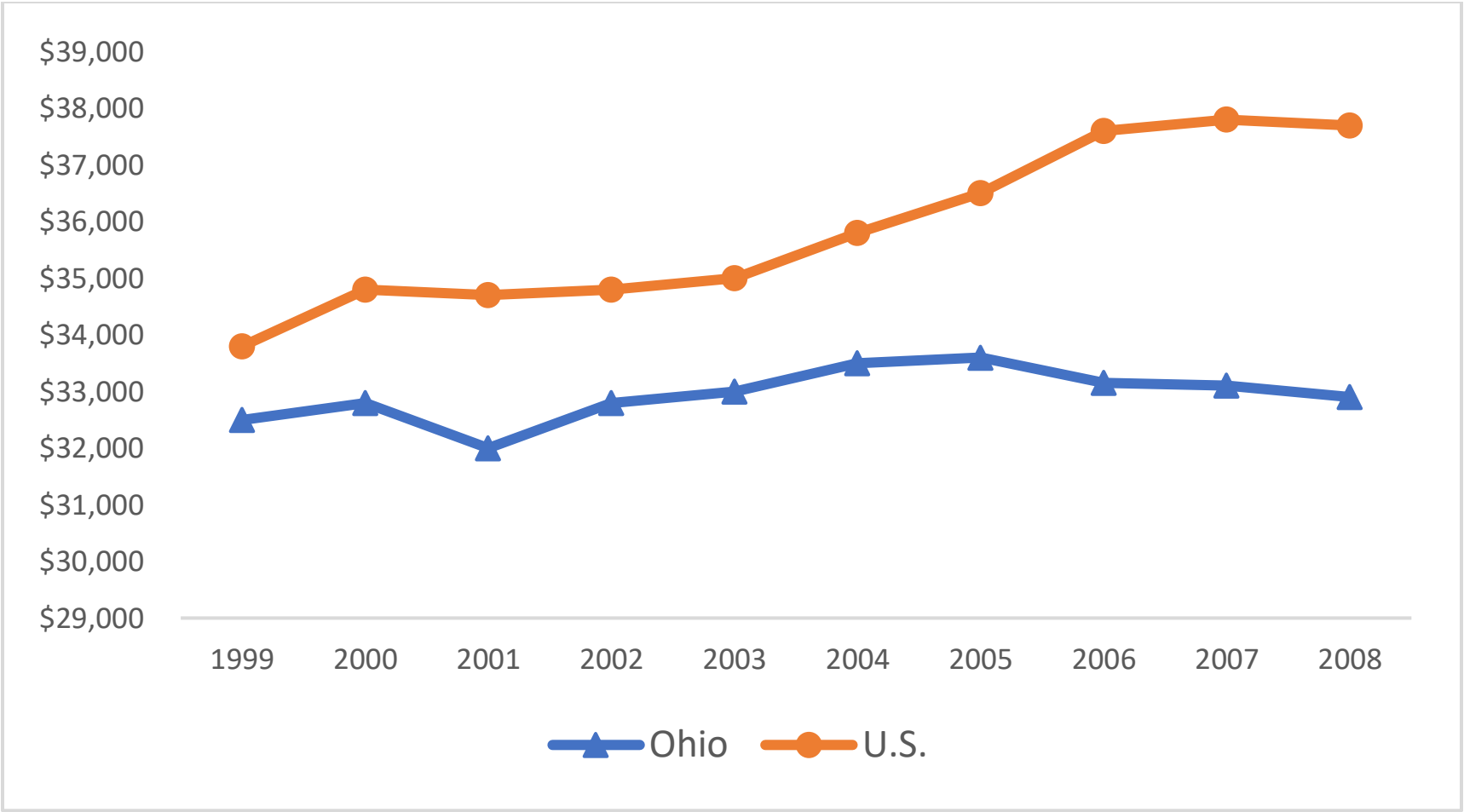

*Figure 8 Per capita real GDP of Ohio residents vs. the U.S. (chained 2000 Dollars)*

In scenario 1, we assume that this financial shock moves a larger fraction of the population to the vulnerable. We also assume that the size of the transfer is 50% more, which means that 22.5% of people move from the LAL to the vulnerable stock. Figure 9 shows the result of the simulation under this assumption from 1995 to 2025.

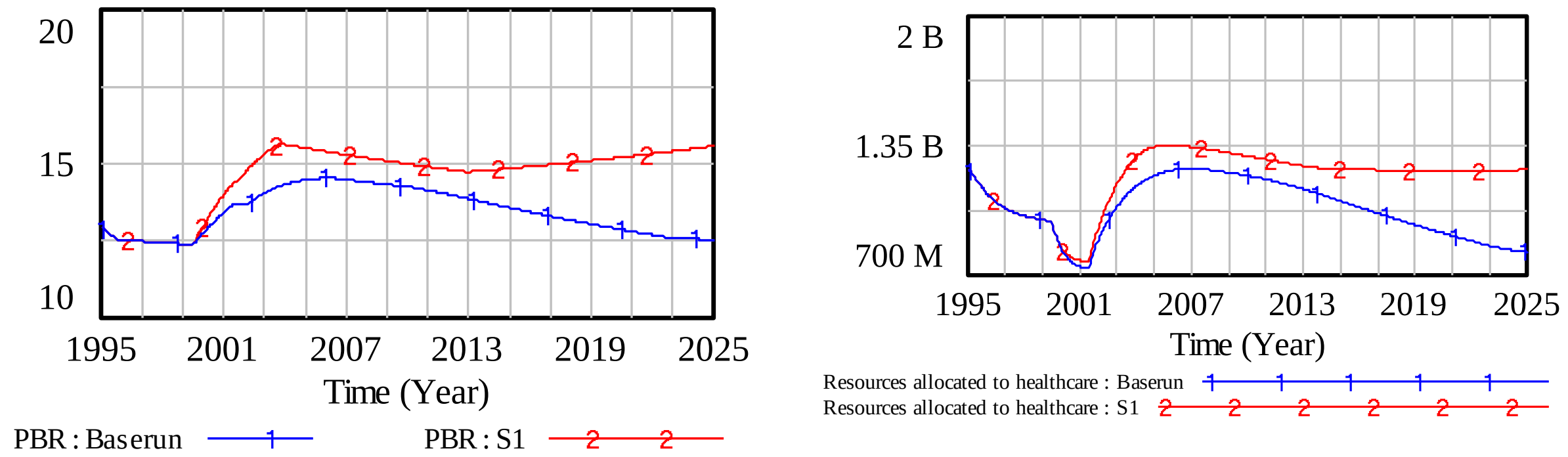

*a. The preterm birth rate in base run vs. scenario 1* *b. Resource allocated to healthcare; base run vs. scenario 1*

*Figure 9 Comparison of base run vs. scenario 1*

Figure 9.a shows that the PBR in the base run has an increasing trend from 2001 until about 2007. But in 2008, it starts to decline and eventually it returns to the rates close to those of 1995. But in the S1, the PBR increases from 2001 until 2003 and the decline starts afterward. The reason for an earlier start in the decline is that the healthcare budget increases faster and it impacts the rate sooner (Figure 9.b). The declining trend continues until 2013. However, after 2013, the PBR starts to increase once again even when there are no other shocks to the system.

The reason for these trends can be better explained by resource allocations. Figure 9.b shows that in both the base run and S1, there is a dip in the resources available, which is due to the financial shock in 2000. This dip leaves fewer resources to be allocated to Medicaid. After the financial shock, more people become eligible for Medicaid and the government needs even more resources for covering the vulnerable. This increase in the healthcare budget leaves fewer resources for categories like education.

Therefore, the resources will be replenished at a lower rate in the long run and the community constantly has to spend more of these scarce resources on healthcare and less on education.

**Scenario 2 (S2)**: Lowering desired preterm birth rate; In this scenario, we assume that the desired rate of preterm birth has a lower value (the lower the PBR, the better). This rate is 11.2 in the base run and we decrease it to nine in the S2.

Figure 10.a shows that the PBR outcomes are better for a few years until 2010 compared to the base run, but the rate of preterm birth crosses the base run after 2011 and starts increasing. Decreasing the desired rate of PBR puts more pressure on the allocation of resources to medical care and constantly directs more funds to healthcare (Figure 10.b). As our community invests more money in healthcare, it drives the resources away from education and other categories that can make the population less vulnerable to undesirable birth outcomes. As a result, the share of the vulnerable increases and there will be less resources to invest in healthcare and other sectors. Therefore, with a few years delay, the outcomes become even worse than before both because of fewer resources and a larger vulnerable population.

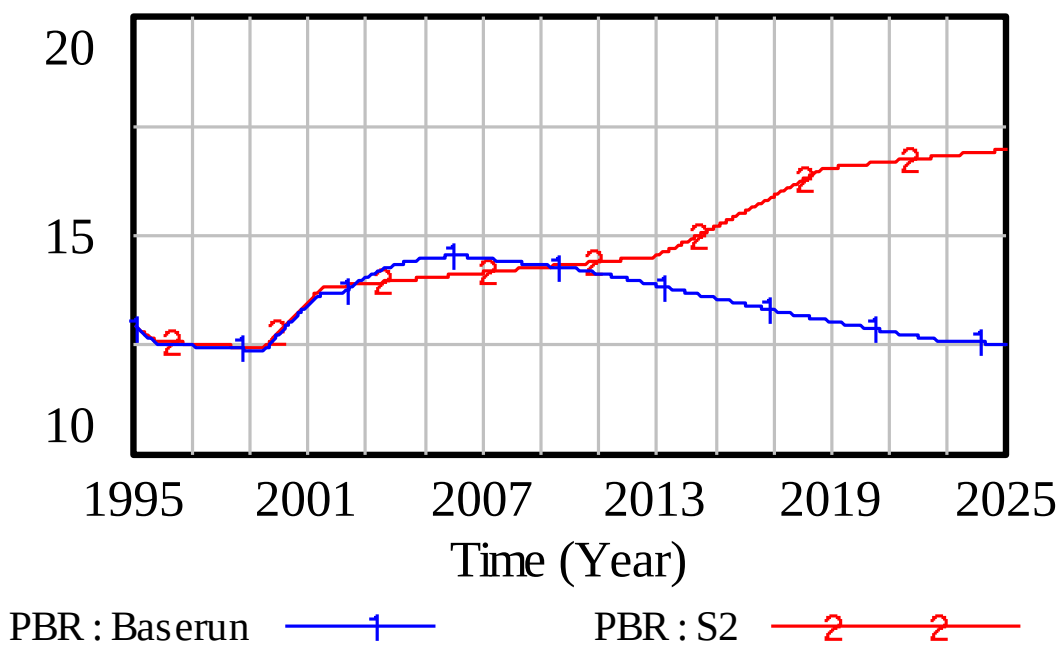

*a. Preterm birth rate in base run vs. scenario 2*

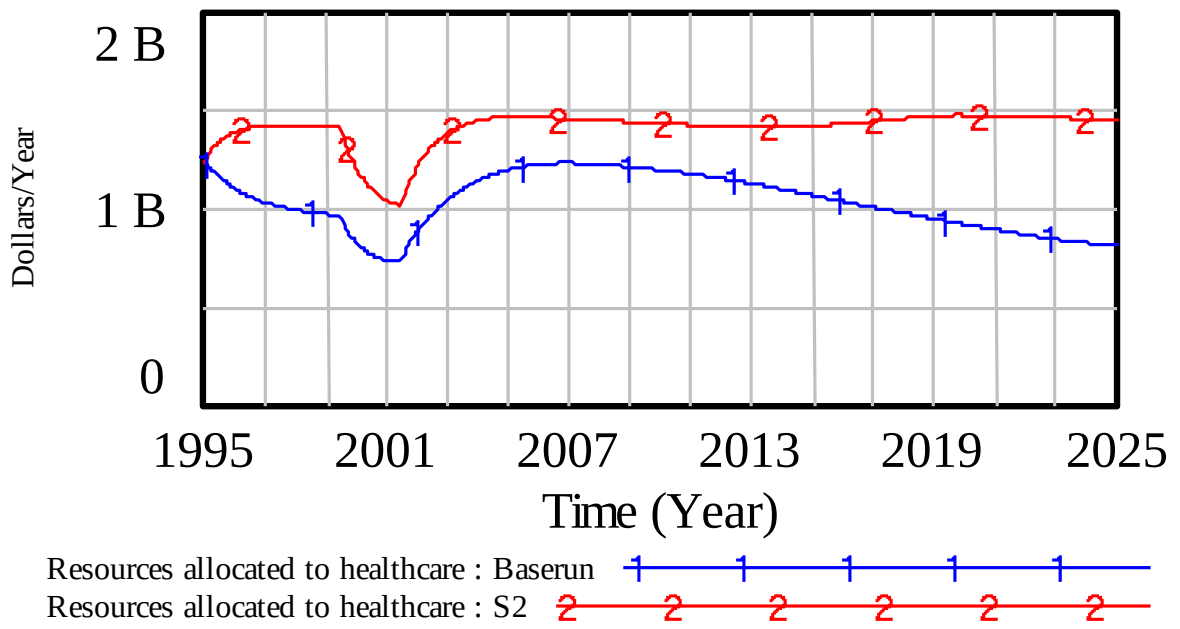

*b. Resources allocated to healthcare; base run vs. scenario 2*

*Figure 10 Comparison of base run vs. scenario 2*

## Discussion

We developed an SD model of preterm births in Cuyahoga County to examine the effect of resource allocation on the long run outcomes of this significant public health problem. Given the dynamic hypotheses of this model, the major insights are: 1) spending more on the expansion of Medicaid cannot solve the problem of high preterm birth rates; 2) it is important to prioritize resource allocation to education and programs that can create sustainable health improvements before spending on expansion of healthcare; 3) aggressive expansion of Medicaid to improve birth outcomes can unexpectedly increase the preterm birth rates in the long-run; and 4) it is necessary to have an optimal resource allocation that can handle the impact of financial shocks because these shocks can both decrease the resources for medical care and increase the vulnerable population.

There is compelling evidence that the allocation of resources was not effective in reduction of preterm birth rates in our case, Cuyahoga County. The governments need to allocate more resources to education in communities with high rates of vulnerable populations in order to improve the social status of people, which consequently improves the health outcomes in the long-run. Despite this need,

Cuyahoga County does not provide the desired quality for public schools. The average performance index of Cuyahoga's schools is 83rd among 88 counties in Ohio. Also, more parents prefer to send their children to private schools. The U.S. Census Bureau estimates that Cleveland has a higher average of private school enrollment at 21.2%, compared to the average of 11.2% for Columbus and the statewide average of 13.1% [49].On the other hand, the number of people who receive Supplemental Nutrition Assistance Program (SNAP) increased from about 200,000 to 243,000 [50]. This increase is despite the decrease in the population of Cuyahoga County from 1.4 to 1.2 million people from 1995 to 2017.

The decision to spend more on medical treatment is one of the few strategies for communities trapped in the vicious cycle of low investment in the key categories. These categories (i.e., education, safety, and social services) can change the direction of the reinforcing loop (i.e., creating a tipping point and turn the vicious cycle into a virtuous cycle) and lead to a healthier and less stressful lifestyle in the long term, subsequently lowering preterm births.

Our findings are not new and they were discussed over 50 years in different research areas that have shown environmental factors such as level of education and smoking make a greater contribution to health outcomes rather than medical expenditure [51, 52]. In particular, the differences in birth outcomes apparently cannot be explained by the differences in access to medical care [31, 53, 54]. The contribution of this paper was framing the preterm birth problem in a causal diagram and building a simulation model. Our model is capable of reproducing the behavior of the PBR and capturing the trends of populations in Cuyahoga County. Another use of this model is testing different desired values for preterm birth and the way that it affects the outcomes of preterm birth over time. Further, one can assess the effects of a financial shock in the community's birth outcomes in the long run.

## Limitations

Considering the broad scope of our study, we were subject to some limitations. The first major limitation is about the ways that populations move to the vulnerable status. We limited the stressors to only financial distress. There are different types of stressors that can increase the allostatic load of people and make them vulnerable to different health problems that we did not consider in our model. These stressors cover a wide range, from institutional racism to the noise level of the neighborhood [12, 55-57]. Some of the statistics for these stressors are provided by a collaboration between the Robert Wood Johnson Foundation and the University of Wisconsin Population Health Institute. However, these statistics are just for recent years and mainly after 2010, even though our simulation start time is 1995. Furthermore, it is not obvious how much of these data series overlap with each other and using them creates the potential for double counting error.

The other limitation of our study is that we considered universal education as the only mechanism for reducing the financial stress of families in the long run. Despite the undeniable effect of education on the prosperity of families and society as a whole, there are other ways to create sustainable improvements. The measures for creating a healthy and sustainable ecosystem include policies and regulations that increase gender equity, long-term food security, access to clean water, and basic sanitation [58].

# Appendix A. Model formulation

## Population sector:

| No | Formula | Unit |
|---|---|---|
| 1 | Initial percent of Vul = 0.28 | Dmnl |
| 2 | Initial county pop = 1.42262e+06 | People |
| 3 | Initial Vul pop = Initial percent of Vul*Initial county pop | People |
| 4 | Initial LAL pop = Initial county pop*(1-Initial percent of Vul) | People |
| 5 | Frac BR LAL = 0.015 | 1/Year |
| 6 | Frac BR Vul = 0.015 | 1/Year |
| 7 | Frac DR LAL = 0.015 | 1/Year |
| 8 | Frac DR Vul = 0.015 | 1/Year |
| 9 | Birth LAL = Frac BR LAL*"Low Allostatic Load (LAL) Population" | People/Year |
| 10 | Vul births = Frac BR Vul*"Vulnerable Population (Vul)" | People/Year |
| 11 | Net Vul flow = ***Frac of Vul immigration due to affordability****"Vulnerable Population (Vul)" | People/Year |
| 12 | LAL death = Frac DR LAL*"Low Allostatic Load (LAL) Population" | People/Year |
| 13 | Vul death = Frac DR Vul*"Vulnerable Population (Vul)" | People/Year |
| 14 | net LAL flow= ***Frac of LAL immigration due to fight or flight response****"Low Allostatic Load (LAL) Population" | People/Year |
| 15 | "Low Allostatic Load (LAL) Population" = INTEG(Birth LAL+Net transition to low-LAL death-net LAL flow-Transition to Vul, Initial LAL pop) | People |
| 16 | "Vulnerable Population (Vul)" = INTEG(net Vul flow+Transition to Vul+Vul births-Net transition to low-Vul death, Initial Vul pop) | People |
| 17 | Transition to Vul = Transition after shock | People/Year |
| 18 | Transition after shock = ***Financial shock****Frac becomming vulnerable*"Low Allostatic Load (LAL) Population" | People/Year |
| 19 | Frac becomming vulnerable = 0.4 | 1/Year |
| 20 | Net transition to low = Upward mobility*"Vulnerable Population (Vul)" | People/Year |
| 21 | Upward mobility = Transition fraction*Family size/Time for education impact*switch of education | 1/Year |
| 22 | Family size = 2 | Dmnl |
| 23 | Time for education impact = 10 | Year |
| 24 | switch of education = 1 | Dmnl |
| 25 | Total pop = "Low Allostatic Load (LAL) Population"+"Vulnerable Population (Vul)" | People |
| 26 | Total births = Vul births+Birth LAL | People |
| 27 | Preterm rate for LAL = 0.104 | Dmnl |
| 28 | LAL preterm births = Birth LAL*Preterm rate for LAL | People/Year |
| 29 | Vul preterm births = Vul preterm OR w prenatal care*Vul births*Preterm rate for LAL | People/Year |
| 30 | Vul preterm OR w prenatal care = (1-switch for medical interventions)*Vul preterm odd ratio +<br>switch for medical interventions*Vul preterm odd ratio*<br>(Medical care effect****Insured frac*** + (1-***Insured frac***) ) | Dmnl |
| 31 | Vul preterm odd ratio = 2.03 | Dmnl |
| 32 | Medical care effect = 0.86 | Dmnl |
| 33 | switch for medical interventions = 1 | Dmnl |
| 34 | Preterm births = Vul preterm births+LAL preterm births | People/Year |
| 35 | PBR = Preterm births/Total births*100 | Dmnl |

## Resource sector:

| No | Formula | Unit |
|---|---|---|
| 1 | Relative contribution of Vul Pop to the resources = 0.58 | Dmnl |
| 2 | Tax contribution of LAL = 3500 | Dollars/(Year*People) |
| 3 | ***Financial shock*** = "Magnitue of financial shock (drop of businesses)"*PULSE( Time of shock , 2 ) | Dmnl |
| 4 | "Magnitue of financial shock (drop of businesses)" = 0.35 | Dmnl |
| 5 | Time of shock = 2000 | Year |
| 6 | Financial resources = ("Vulnerable Population (Vul)"*Relative contribution of Vul Pop to the resources+"Low Allostatic Load (LAL) Population")*Tax contribution of LAL*( 1-Financial shock) | Dollars/Year |
| 7 | Resources = INTEG(Income from taxes-Other-Resources allocated to healthcare-Resources on schools, 4e+09) | Dollars |
| 8 | Other = Frac of resources on other*Resources | Dollars/Year |
| 9 | Frac of resources on other = 0.4 | 1/Year |
| 10 | Resources on schools = (1-Percentage of resources allocated to Medicaid-Frac of resources on other)*Resources | Dollars/Year |
| 11 | Resources allocated to healthcare = Percentage of resources allocated to Medicaid*Resources | Dollars/Year |
| 12 | Percentage of resources allocated to Medicaid = Gap and pressure on resource allocation(Realized gap) | 1/Year |
| 13 | Gap and pressure on resource allocation = [(-10,0)-(15,0.7)],(-10,0.101316),(-4,0.11),(0.168196,0.17193),(2.76758,0.297807),(4.90826,0.389912),(7.43119,0.439035),(10.1835,0.475877),(14.5,0.482018),(15.0765,0.488158) | 1/Year |
| 14 | Realized gap = DELAY1I( Gap, Time to realize the gap by policymakers , 3) | Dmnl |
| 15 | Gap = PBR-Desired PBR rate | Dmnl |
| 16 | Avg medical cost of insurance = 4300 | Dollars/(People*Year) |
| 17 | Desired medical resources = "Vulnerable Population (Vul)"*Avg medical cost of insurance | Dollars/Year |
| 18 | Adequacy of resources for insurances = Resources allocated to healthcare*Federal Match-Desired medical resources | Dollars/Year |
| 19 | Federal Match = 1.75 | Dmnl |
| 20 | Changes in insurances availability = Adequacy of resources for insurances/Avg medical cost of insurance | People/Year |
| 21 | Insurances = INTEG(Changes in insurances availability, 500000) | People |
| 22 | ***Insured frac*** = min(max(DELAY1I( Insurances/"Vulnerable Population (Vul)",Time to implement policies, 0.5),0),1) | Dmnl |
| 23 | Time to implement policies = 1 | Year |
| 24 | School age percentage = 0.16*2 | 1/Year |
| 25 | Vul school age children = School age percentage*"Vulnerable Population (Vul)" | People/Year |
| 26 | LAL school age children = School age percentage*"Low Allostatic Load (LAL) Population" | People/Year |
| 27 | School age children = LAL school age children+Vul school age children | People/Year |
| 28 | Desired frac of school funding = 0.65 | Dmnl |
| 29 | Desired number of school funds = School age children*Desired frac of school funding | People/Year |
| 30 | Local government match = 2.3 | Dmnl |
| 31 | Avg cost of schooling = 13000 | Dollars/People |
| 32 | Number of school funds available = (Resources on schools*Local government match)/Avg cost of schooling | People/Year |
| 33 | Adequacy of school funds = Number of school funds available-Desired number of school funds | People/Year |
| 34 | School funds status = INTEG(Adequacy of school funds, 0) | People |
| 35 | Vul frac = "Vulnerable Population (Vul)"/("Low Allostatic Load (LAL) Population"+"Vulnerable Population (Vul)") | Dmnl |
| 36 | ***Transition fraction*** = School age percentage* IF THEN ELSE(School funds status>0, 1,max(School funds status/LAL school age children,-1)*(Desired frac of school funding-Vul frac) | Dmnl |

## Crime sector:

| No | Formula | Unit |
|---|---|---|
| 1 | Crime rate of LAL = 450/100000 | Crimes/Year |
| 2 | Relative crime rate of the Vul pop = 4 | Dmnl |
| 3 | Crime rate of the community = (("Low Allostatic Load (LAL) Population"+"Vulnerable Population (Vul)"*Relative crime rate of the Vul pop)*Crime rate of LAL)/("Low Allostatic Load (LAL) Population" +"Vulnerable Population (Vul)")*100000 | Crimes/Year |
| 4 | National violent crime rate = [(1995,0)-(2017,900)],(1995,684.46),(1996,636.64),(1997,611),(1998,567.6),(1999,523),(2000,506.5),(2002,494.4),(2003,475.8),(2004,463.2),(2005,469),(2006,479.3),(2007,471.8),(2008,458.6),(2009,431.9),(2010,404.5),(2011,387.1),(2012,387.8),(2013,369.1),(2014,361.6),(2015,373.7),(2016,386.3),(2017,382.9),(20015,504.5) | Crimes/Year |
| 5 | Relative crime = Crime rate of the community/National violent crime rate(Time) | Dmnl |
| 6 | Delay for people to receive crime info = 1 | Year |
| 7 | Perception of crime = smooth(Relative crime,Delay for people to receive crime info) | Dmnl |
| 8 | Crime perception and immigration = [(0.6,-0.03)-(2,0.1)],(0.66,-0.015),(0.8,-0.01),(0.9,-0.005),(1,0),(1.1,0.005),(1.25,0.01),(1.5,0.015),(2,0.015) | 1/Year |
| 9 | switch for outmigration = 1 | Dmnl |
| 10 | Frac of LAL immigration due to fight or flight response = Crime perception and immigration(Perception of crime)*switch for outmigration | 1/Year |
| 11 | switch for immigration = 1 | Dmnl |
| 12 | Time delay = 1 | Year |
| 13 | "Relative Vul immigration (net Vul migration)" = 0.45 | Dmnl |
| 14 | Frac of Vul immigration due to affordability = smooth( Frac of LAL immigration due to fight or flight response *"Relative Vul immigration (net Vul migration)"*switch for immigration, Time delay) | 1/Year |
| 15 | net migration = -Frac of LAL immigration due to fight or flight response*"Low Allostatic Load (LAL) Population" + Frac of Vul immigration due to affordability*"Vulnerable Population (Vul)" | People/Year |